\documentclass[journal=jacsat,manuscript=article]{achemso}

\usepackage[version=3]{mhchem} % Formula subscripts using \ce{}
\usepackage{algorithm}
 \usepackage{graphicx}
 \usepackage{subcaption}
 \usepackage{pdfpages}
\usepackage{algpseudocode} 
\usepackage{siunitx}
\usepackage[section]{placeins}
\author{Pier Paolo Poier}
\affiliation{Sorbonne Universit\'e, LCT, UMR 7616 CNRS, Paris, France}
\email{pier.poier@sorbonne-universite.fr}
\author{Théo Jaffrelot Inizan}
\affiliation{Sorbonne Universit\'e, LCT, UMR 7616 CNRS, Paris, France}
\author{Olivier Adjoua}
\affiliation{Sorbonne Universit\'e, LCT, UMR 7616 CNRS, Paris, France}
\author{Louis Lagardère}
\affiliation{Sorbonne Universit\'e, LCT, UMR 7616 CNRS, Paris, France}
\alsoaffiliation{Sorbonne Universit\'e, IP2CT, FR 2622 CNRS, Paris, France}
\author{Jean-Philip Piquemal}
\affiliation{Sorbonne Universit\'e, LCT, UMR 7616 CNRS, Paris, France}
\alsoaffiliation{The University of Texas at Austin, Department of Biomedical Engineering, TX, USA}
\email{jean-philip.piquemal@sorbonne-universite.fr}

\title[An \textsf{achemso} demo]
  {Accurate Deep Learning-aided Density-free Strategy for Many-Body Dispersion-corrected Density Functional Theory}

\abbreviations{IR,NMR,UV}
\keywords{American Chemical Society, \LaTeX}

\begin{document}


\begin{abstract}
Using a Deep Neuronal Network model (DNN) trained on the large ANI-1 data set of small organic molecules, we propose a transferable density-free many-body dispersion model (DNN-MBD). The DNN strategy bypasses the explicit Hirshfeld partitioning of the Kohn-Sham electron density required by MBD models to obtain the atom-in-molecules volumes used by the Tkatchenko-Scheffler polarizability rescaling. The resulting DNN-MBD model is trained with minimal basis iterative Stockholder atomic volumes and, coupled to Density Functional Theory (DFT), exhibits comparable (if not greater) accuracy to other approaches based on different partitioning schemes. Implemented in the Tinker-HP package, the DNN-MBD model decreases the overall computational cost compared to MBD models where the explicit density partitioning is performed. Its coupling with the recently introduced Stochastic formulation of the MBD equations (\textit{J. Chem. Theory Comput.}, \textbf{2022}, \textit{18}, 3, 1633–1645) enables large routine dispersion-corrected DFT calculations at preserved accuracy. Furthermore, the DNN electron density-free features extend MBD's applicability beyond electronic structure theory within methodologies such as force fields and neural networks.

%While the electron density-free features of the SDNN-MBD model further decreases the computational cost enabling routine large dispersion-corrected DFT computations, it also extends its applicability beyond electronic structure theory, opening a route for its use within force fields and neural networks.
%\begin{tocentry}

%Some journals require a graphical entry for the Table of Contents.
%This should be laid out ``print ready'' so that the sizing of the

%Inside the \texttt{tocentry} environment, the font used is Helvetica
%8\,pt, as required by \emph{Journal of the American Chemical

\begin{figure}[!htb]
\centering
   \includegraphics[width=0.5\linewidth]{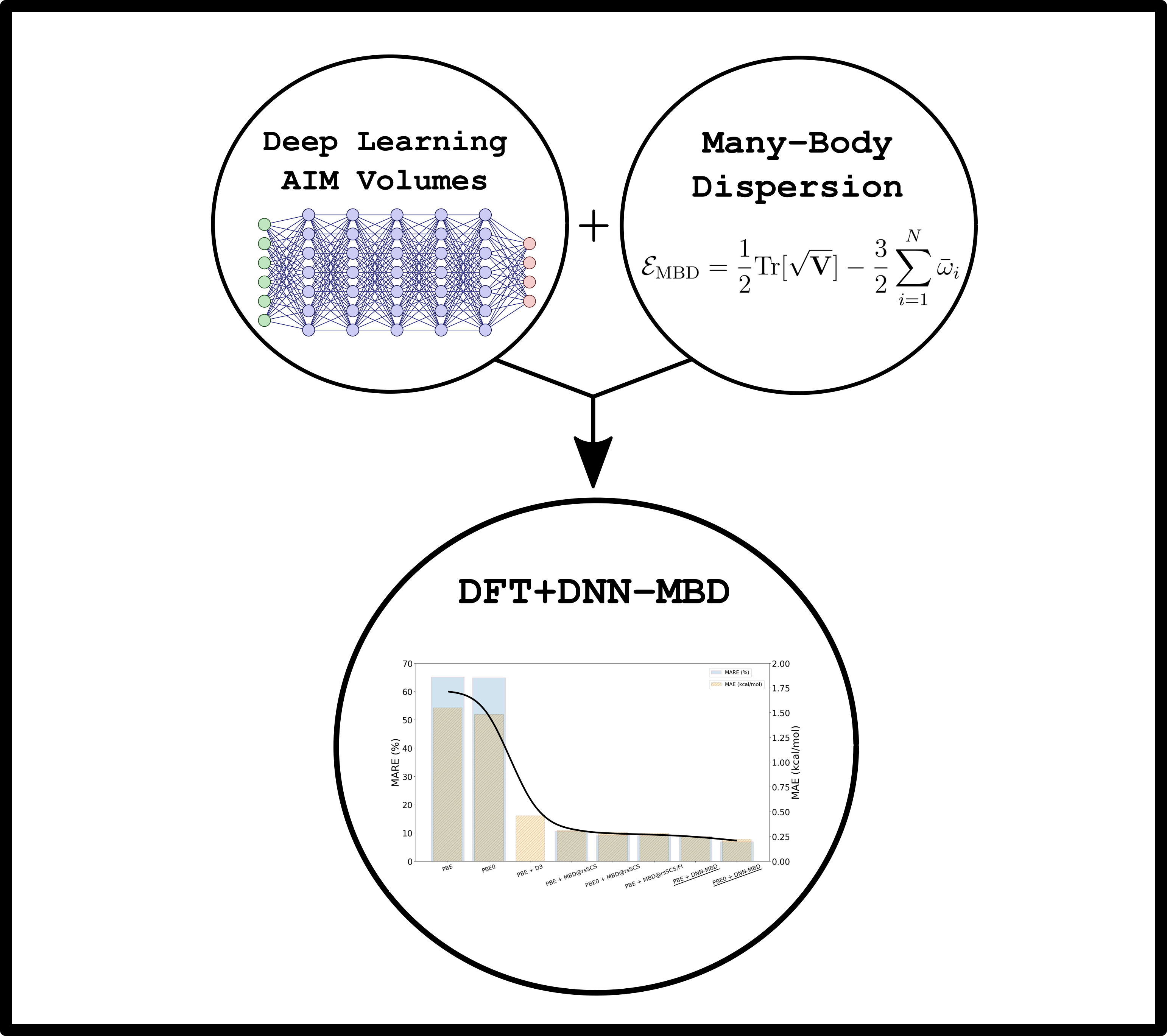}
   %\caption{Convergence of the simple fixed-point rs-SCS polarizabilities for ten different frequencies (nodes in the GL quadrature) as function of the number of the iterations. The convergence is followed via the logarithm of the residual $R$ defined in Eq.\eqref{resid}.}
   \label{fig:nodiis} 
\end{figure}
%The surrounding frame is 9\,cm by 3.5\,cm, which is the maximum
%permitted for  \emph{Journal of the American Chemical Society}
%graphical table of content entries. The box will not resize if the
%content is too big: instead it will overflow the edge of the box.

%This box and the associated title will always be printed on a
%separate page at the end of the document.

%\end{tocentry}

\end{abstract}

%\section{Introduction}
Since its original formulation in 1965, Kohn-Sham Density Functional Theory\cite{kohnsham} (KS-DFT) has become the most popular family of electronic structure methods. KS-DFT represents in fact the cheapest way for introducing electronic correlation as its computational cost is similar to that of the Hartree-Fock method. KS-DFT is based on the idea of evaluating the kinetic energy from a Slater determinant thus assuming the electrons to be non-interacting. This apparently crude assumption  actually leads to big improvements in describing chemical bonding compared to, for example, the use of the Thomas-Fermi kinetic energy formulation. The difference between the Slater determinant kinetic energy representation and the true one, together with the difference between the true total electronic interaction and the exchange energies represents, in KS-DFT, the key contribution to the exchange-correlation functional which remains, however, unknown.\\
In practice, the plethora of existing KS-DFT variants differentiate themselves in the way the exchange-correlation functional is approximated. Typically it is assumed to be a functional of the local electron density and eventually of its gradient and Laplacian. As a consequence, only local contributions to electronic correlation are included and this explains the general inadequacy of DFT methods to describe dispersion interactions which, on the other hand, have roots in long-range electronic correlation.\\
To retain the pleasant computational performances of KS-DFT methods, several dispersion corrections have been proposed.\cite{grimmechemrev} Among these, the popular and successful approach of Grimme includes dispersion via empirical pairwise $C_6$ terms.\cite{grimme,grimme2,grimme3} This is particularly appealing in virtue of its nearly zero additional computational cost.\\
A further approach is to replace the empirical pairwise terms with ones obtained from quantities coupled to the molecular electron density. For example, in Becke and Johnson's model, pairwise $C_6$ coefficients are written in terms of atomic polarizabilities and the averaged exchange-hole dipoles corresponding to each of the two atoms in the pair.\cite{beckejohnson,johnson} In the alternative approach proposed by Tkatchenko and Scheffler (TS),\cite{tksche} pairwise $C_6$ coefficients are instead expressed in terms of accurate free atom reference data as well as atoms-in-molecule (AIM) polarizabilities obtained from the rescaling of the corresponding AIM volumes computed via the Hirshfeld partitioning of the molecular electron density.\cite{hirsh} \\
One limitation of the above mentioned pairwise approaches is the impossibility of capturing non-additive many-body dispersion (MBD) effects, which inclusion has recently been shown important in modelling extended systems, supramolecular complexes and proteins in solutions, among others.\cite{mbd_extended,mbd_qmc,plasmonic1,plasmonicmbd}\\
The non-additive long-range character of dispersion interactions has been modeled via a set of coupled fluctuating dipoles \cite{mbd_cfd,mbd_cfd2} (CFD) or alternatively by quantum Drude oscillators.\cite{qdojordan,qdo,qdomartyna,qdojordan2}\\ 
In recent years, Tkatchenko, DiStasio Jr., Ambrosetti et al. have proposed a range-separated many-body dispersion model based on the CFD where the self-consistent screening  of a set of atomic polarizabilities is performed (MBD@rsSCS).\cite{scs,mbd_rsscs} The MBD@rsSCS model is appealing not only for introducing non-additive many-body dispersion effects but also since it relies, \emph{de facto}, on a single range-separation parameter which is tuned according to the choice of the exchange-correlation functional employed.\\ The MBD@rsSCS keeps in fact the spirit of the TS approach where AIM polarizabilities and van der Waals radii are obtained via the Hirshfeld partitioning of the density.\\
The Hirshfeld method leads to AIM densities which minimize the Kullback-Lieber divergence corresponding to the information loss upon molecule formation where this solid mathematical condition is used as a basis for the development of new information-theoretic partitioning methods.\cite{inftheorypart}\\
As discussed in references \cite{ayers1,ayers2} Hirshfeld partitioning makes its resulting AIM densities as close as possible to the ones of the isolated atoms, consequently AIM's properties turn out to be as similar as possible to those of the free atoms. This is particularly evident in the magnitude of Hirshfeld atomic charges, being too small in magnitude for reproducing the molecular electrostatic potential (ESP) or in modeling AIM polarizabilities in ionic and covalent crystals where the Hirshfeld partitioning leads to unrealistically large polarizabilities of cations which can even be found larger than those of the anions.\cite{bucko_tshi}\\
The above mentioned shortcomings were ameliorated by the Iterative Hirshfeld (HI) scheme\cite{hirsh_it} where the reference atomic density employed in the partitioning is constructed as a linear combination of the two densities relative to the atomic oxidation states closest to the fractional number of electrons assigned by the partitioning at a given iteration.\\
The ESP computed from HI atomic charges have proven to agree remarkably well with \emph{ab initio} computed reference.\cite{esp_hi} In addition, the use of HI derived AIM polarizabilities leads to more realistic dispersion coefficients\cite{bucko_tshi} especially in ionic systems and adsorption phenomena on surfaces of ionic solids where the HI scheme used within the TS dispersion model improves interaction energies.\cite{bucko_jctc} HI partitioning has also been employed in the MBD@rsSCS model replacing the original Hirshfeld scheme \cite{hi_mbd} and its use in in connection to the fractionally ionic AIM polarizabilities leads, in the just mentioned challenging systems, to reduced errors.\cite{fracpol}\\
Despite the improvements carried by the HI partitioning, the scheme remains affected by a shortcoming arising from the density interpolation for negatively charged atoms as this procedure is, for some species, ill-defined. This arises from the fact that free anions such as $\ce{N-}$ and \ce{O ^{2-}} (or in general any doubly negative ion) are not bound and their reference electron densities, computed at a complete basis set (CBS), result in a detached electron.\\
The iterative Stockholder atom (ISA) scheme, on the other hand, is not affected by this problem as the partitioning does not require reference atomic densities computed from isolated atoms at different ionic states as they are rather obtained from a spherical averaging of the molecular density using nuclei as expansion points.\cite{isa,ISAAlston} The minimal basis iterative Stockholder atom (MBISA), a variant of the ISA method, have proven successful in the atomic polarizability rescaling approach employed by the TS scheme as well as in reproducing \emph{ab initio} ESP from atomic point charges\cite{mbisa} and for this reasons its use in connection to the MBD@rsSCS model is particularly appealing.\\ 
AIM properties are local quantities which depend on the near chemical environment and thus carry a certain degree of transferability. In particular, the TS polarizability rescaling scheme (employed in the MBD@rsSCS) makes use of AIM volumes which are well suited to be computed via deep neural network (DNN) where the environment vector associated to an atom's surrounding is defined within a local cutoff. The potential of deep learning in capturing local atomic properties has been proved by Isayev and co-workers whose multi-output DNN model successfully predicts AIM properties ranging from multipoles to volumes.\cite{aimnet} \\

In this Letter we present a hybrid DNN-aided MBD@rsSCS model (DNN-MBD) where the AIM volumes ratio employed in the TS polarizability rescaling are generated by a deep neural network trained on the ANI-1 data set (approximately 4.6 million structures) containing MBISA AIM volumes.\cite{ANI1dataset}\\
For the common S66x8 benchmark set\cite{s66x8}, the DNN-MBD model coupled to the common PBE/PBE0 density functionals, exhibits excellent interaction energies  while completely bypassing the electron density partitioning with a consequent computational cost reduction. This electron density-free DNN-MBD approach employed in connection to our recently proposed linear scaling stochastic MBD@rsSCS formulation \cite{mbd_stoch}, allows for modelling non-additive long-range dispersion interactions of up-to-millions atom systems at a very low computational cost without compromising the accuracy.\\
We note that kernel-ridge regression approaches to model AIM polarizabilities have been proposed in modelling dispersion interactions.\cite{bereau_vdw, mldisp} This approach, however, is characterized by a $\mathcal{O}(N^2)$ and $\mathcal{O}(N^3)$ scaling of the required memory and computational cost involved in the model's training respectively, $N$ being the size of the data set. These non-linear scaling prevents the applicability of kernel-ridge approaches on very large and diverse data sets, necessary for the generation of general-purpose MBD models. Additionally, the poor scaling with the number of processes limits its use on large systems. Here instead we generalize the approach to model MBD interactions to a much broader class of systems thanks to the employed model's flexibility and broad data set, without affecting the model's accuracy and linear scalability. \\

We will, in the following, proceed by briefly recalling the key concepts of the standard MBD@rsSCS model before introducing the DNN-MBD hybrid model and its performances.\\

\begin{figure}[!htb]
\centering
   \includegraphics[width=0.9\linewidth]{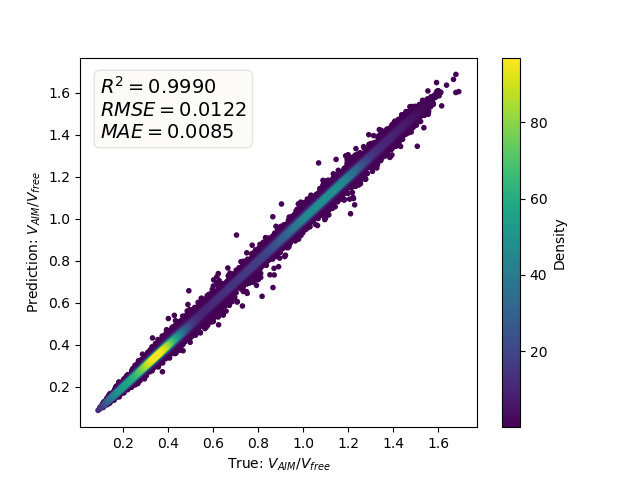}
   \caption{Atomic volume correlation plot comparing the DNN prediction to DFT reference calculations for 1/100 of the validation set. The color bar scale reflects the density of points and correlate with the atomic volumes ratio distribution (Figure 2 of the SI).}
   \label{fig:mlpred} 
\end{figure}

As a starting point in this discussion, we examine the TS polarizability rescaling in Eq.\eqref{tspol}, where $\alpha_i$ and $V_i$ represent the TS static polarizability and AIM volume respectively of the i-th atom while the zero superscript denotes free atom reference quantities. 
\begin{equation}\label{tspol}
  \alpha_i=\biggl(\frac{V_i}{V_i^0}\biggr)\alpha_i^0  
\end{equation}
The AIM volume $V_i$ is obtained by solving the integral in Eq.\eqref{vaim} where $\rho(\mathbf{r})$ is the Kohn-Sham molecular electron density which, via the partitioning-specific weight function $w_i(\mathbf{r})$, is decomposed into its AIM densities $\{\rho_i(\mathbf{r})\}$. 
\begin{equation}\label{vaim}
\begin{split}
    V_i&=\int \mathbf{r}^3\rho_i(\mathbf{r}) d^3\mathbf{r}\\
    \rho_i(\mathbf{r})&=w_i(\mathbf{r})\rho(\mathbf{r})
\end{split}
\end{equation}
Once the set of static AIM polarizabilities in Eq.\eqref{tspol} is obtained, a correspondent set of frequency-dependent ones is generated via Eq.\eqref{frqpol}, where this time $\omega_j^0$ and  $C_{6,j}^0$ are the free atom characteristic excitation frequency and first dispersion coefficient. 
\begin{equation}
    \label{frqpol}
    \begin{split}
    \alpha_j(i\nu)&=\frac{\alpha_j}{1-(i\nu/\omega_j^0)^2}\\
    \omega_j^0&=\frac{4}{3}\frac{C^0_{6,j}}{\bigr(\alpha_j^0\bigr)^2}
       \end{split}
\end{equation}
These frequency-dependent polarizabilities are, in the MBD@rsSCS model, gathered as diagonal elements of the frequency-dependent superpolarizability matrix $\mathbf{A}(i\nu)$ being one of the entries in the Dyson-like equation below which solution provides the screened super-polarizability matrix $ \Bar{\mathbf{A}}(i\nu)$.
\begin{equation}
    \label{dyson}
    \Bar{\mathbf{A}}(i\nu)=\mathbf{A}(i\nu)-\mathbf{A}(i\nu)\mathbf{T}^\text{SR}(i\nu)\Bar{\mathbf{A}}(i\nu)
\end{equation}
The $\mathbf{T}^\text{SR}$ represents a damped dipole-dipole interaction operator applied to the Coulombic interaction of two frequency-dependent spherical Gaussian charge distributions where its explicit expression, together with the one for $\mathbf{A}(i\nu)$, can be found in reference.\cite{mbd_stoch} We note here that the Fermi damping function employed in the definition of $\mathbf{T}^\text{SR}$ makes use of AIM van der Waals radii which can also be obtained by a volume rescaling similarly to what discussed for polarizabilities.\cite{tksche}\\
The solution of Eq.\eqref{dyson} for a set of frequencies, and a consequent partial contraction of the converged $\{\Bar{\mathbf{A}}(i\nu)\}$, gives a set of screened frequency-dependent atomic polarizabilities $\{\Bar{\alpha}_j(i\nu)\}$ which are used to approximate the Casimir-Polder integral providing screened characteristic excitation frequencies $\{\Bar{\omega}_j\}$.
\begin{equation}
    \label{wbar}
    \begin{split}
    &\Bar{C}_{6,j}=\frac{3}{\pi}\int_{0}^\infty \Bar{\alpha}_j(i\nu)\Bar{\alpha}_j(i\nu)d\nu\\
   &\Bar{\omega}_j=\frac{4}{3}\frac{\Bar{C}_{6,j}}{\bigr[\Bar{\alpha}_j(0)\bigr]^2}
     \end{split}
\end{equation}
The set of screened excitation frequencies as well as the screened static atomic polarizabilities define the MBD potential matrix shown in Eq.\eqref{vmbd} for a general $ij$ block. $\mathbf{T}^\text{LR}$ represents the range-separated damped dipole-dipole interaction matrix which explicit expression is also found in reference.\cite{mbd_stoch} 
\begin{equation}
    \label{vmbd}
    \mathbf{V}_{ij}=\delta_{ij}\Bar{\omega}_i^2+(1-\delta_{ij})\Bar{\omega}_i\Bar{\omega}_j\sqrt{\Bar{\alpha}_i(0)\Bar{\alpha}_j(0)} \mathbf{T}^\text{LR}_{ij} 
\end{equation}
The trace of $\sqrt{\mathbf{V}}$ defines the interaction energy $\mathcal{E}_\text{int}$ of the CFDs in the system\cite{mbd_stoch} while its zero-point value $\mathcal{E}_0$ is given by the sum of all screened excitation frequencies. Finally the difference between $\mathcal{E}_\text{int}$ and $\mathcal{E}_0$ gives the target MBD@rsSCS energy, Eq.\eqref{embd}, which is coupled to the KS-DFT one to include non-additive dispersion contributions.
\begin{equation}
    \label{embd}
    \mathcal{E}_\text{MBD}=\mathcal{E}_\text{int}-\mathcal{E}_\text{0}=\frac{1}{2}\text{Tr}[\sqrt{\mathbf{V}}]-\frac{3}{2}\sum_{i=1}^N\Bar{\omega}_i
\end{equation}
In the original MBD@rsSCS model just briefly reviewed, $\mathcal{E}_\text{MBD}$ is coupled to the molecular electron density via AIM volume partitioning introduced in Eq.\eqref{vaim}.\\
In this Letter instead we show that the explicit electron density partitioning can be avoided by learning AIM volumes via a DNN model without affecting the original MBD@rsSCS model's accuracy.\\
Bereau et al. and more recently, Mulhi et al. used ML on atomic volumes inside vdW model to capture many body effects. \cite{bereau_vdw, mldisp} Both have developed a Gaussian approximation potential (GAP) force field on TS polarizability rescaling. While GAP has shown to outperform neural networks in predicting energies with small-sized data set, e.g few thousands of data, its poor computational scaling $\mathcal{O}(N^3)$ prevents its use on very large training sets and thus to build a general purpose MBD model \cite{performanceMLP} Finally, these models are either restricted to pairwise interactions or do not scale linearly with respect to the number of atoms as our Stochastic reformulation of the MBD equations was introduced only recently.\cite{mbd_stoch}\\
Isayev et al. \cite{aimnet} recently extended their 5 million chemical conformations, the ANI-1 data set, with atomic volumes computed at the $\omega$B97x/def2-TZVPP level with MBISA partitioning. In virtue of its size and diversity, this data set is here employed in building our DNN to be coupled to the MBD@rsSCS model. Here we restrict ourselves to structure composed of only C, H, N and O, thus reducing the actual data set size to 4.6 million conformations.\\
In the MBISA weight function $w_i(\mathbf{r})$,  each of the reference pro-atomic densities $\rho^0_i(\mathbf{r})$ is expanded into $m_i$ Slater functions, $m_{i}$ being the number of
shells of atom i placed at $\mathbf{R}_i$.
\begin{equation}\label{mbisa}
\begin{split}
    w_i(\mathbf{r})=&\frac{\rho^0_i(\mathbf{r})}{\sum_{j=1}^N\rho^0_j(\mathbf{r})}\\
\rho^0_i(\mathbf{r})=&\sum_{\sigma=1}^{m_{i}}\frac{N_{i,\sigma}}{k^3_{i,\sigma}8\pi}\exp{\biggl(-\frac{\| \mathbf{r}-\mathbf{R}_i\| }{k_{i,\sigma}}\biggr)}
\end{split}
\end{equation} 
In the scheme, the population $N_{i,\sigma}$ and width $k_{i,\sigma}$ of each shell are free-variables which are optimized so that the loss of information upon molecule formation is minimized.\cite{mbisa}\\To handle such large data set, a deep neural network is the natural choice. \cite{NNvdw_material} In particular, we use as machine learning model a feed-forward DNN with the ANI-like symmetry functions (SFs)\cite{ANI1x}. The ANI's SFs are a subfamily of Behler-Parinello's ones\cite{behler-parrinello} which traduce an atomic local environment $i$ into an atomic environment vector (AEV) $G_{i} = \{G^{R}_{i}, G^{A}_{i}\}$ where $G^{R}_{i}$ and $G^{A}_{i}$ represent its radial and angular contributions respectively. Although SFs development is an intensive field of research and more accurate models have been developed since ($\omega$ACSF\cite{wACSF}, SOAP\cite{SOAP} among others), we stick to the ANI's original SFs as they were shown to successfully predict complex local properties such as, in the case of AIMNET, multipoles and volumes.\cite{aimnet} Moreover, ANI's SFs have the great advantage of being computationally efficient as they rely on 2-body terms thus making the overall DNN model linear scaling with the system's size.\\
The DNN part of the combined DNN-MBD model relies on Scikit-learn\cite{scikit}, PyTorch \cite{pytorch} and TorchAni\cite{torchani}. They are all included in the Tinker-HP neural network module, which implementation will be detailed in a forthcoming dedicated paper (T. Jaffrelot Inizan et al., 2022).\\  
We kept the original ANI's SFs parameters %(, $\zeta$, $\eta$, $R_{s}$)
as we did not see major differences after tuning them. We empirically tested multiple neural network architectures (further details are found in the Supporting Information (SI) Figure 3) and the best performance was obtained with 5 hidden layers . The atomic element's neural network architectures are H 160:128:96:48:1; C 144:112:96:48:1; N 128:112:96:48:1; O 128:112:96:48:1. We observed that by adding 1 extra layer to the original ANI-1x model architecture slightly increases the performance of the model while making it more flexible. Indeed, in the original ANI-1x model, the last layer is composed of 96 neurons, and adding an extra 48 neurons layer may prevent loss of information. We used the Exponential Linear Units (ELU) activation function\cite{DNNELU} while the model's parameters were initialized with the so-called ``He'' initialization and updated with Hutter's AdamW algorithm during the training procedure.\cite{AdamW} Within the AdamW algorithm, the factor was set to 0.5 and the patience to 100. The initial learning rate was set to $10^{-3}$ and the early stopping learning rate was set to $10^{-6}$. The ANI-1 dataset was shuffled and split into training and validation set containing 80$\%$ and 20$\%$ respectively of the full dataset. The networks were trained for 6000 epochs with a batch size of 2560.\\ %(PIER: we move this to SI since we do not have a graph and this may be confusing without one looking at a figure....)Only the AIMNet-like model was stopped before the end as its learning rate reach the early stopping limit within 3000 epochs. \\
The ANI-1 data set, upon which our DNN model is trained, consists of AIM volumes computed at the $\omega$B97x/def2-TZVPP level. The model is trained on volume ratios rather than pure AIM volumes as the narrower distribution of the former allows for a DNN's better performance without the need for re-scaling. Indeed, the atomic volumes ratio for C, H, O, N (see Figure 1 of the SI) is between 0.1 and 1.6. Free atom volumes are computed at the same level as AIM ones. The correlation plots between the DNN model and the \emph{ab initio} validation set reference is depicted in Figure \ref{fig:mlpred}. The root-mean-square-error (RMSE) and mean-absolute-error (MAE) are respectively 0.012 and 0.008 which is much less than the smallest value of the data set showing the good accuracy of our model.
% Overall, our model accuracy is comparable to AIMNet on the direct volumes prediction, as they found an RMSD of 0.03$\AA^3$ on the COMP6-SFCl benchmark. Additionally, the model is extremely efficient as it can be use on GPU while avoiding the SCF-like procedure of the multi-output AIMNet network. 
The final DNN model and the dataset used for the training can be download directly via the Zenodo repository located at the following address \cite{zenodo}. \\

\begin{figure}[!htb]
\centering
   \includegraphics[width=0.9\linewidth]{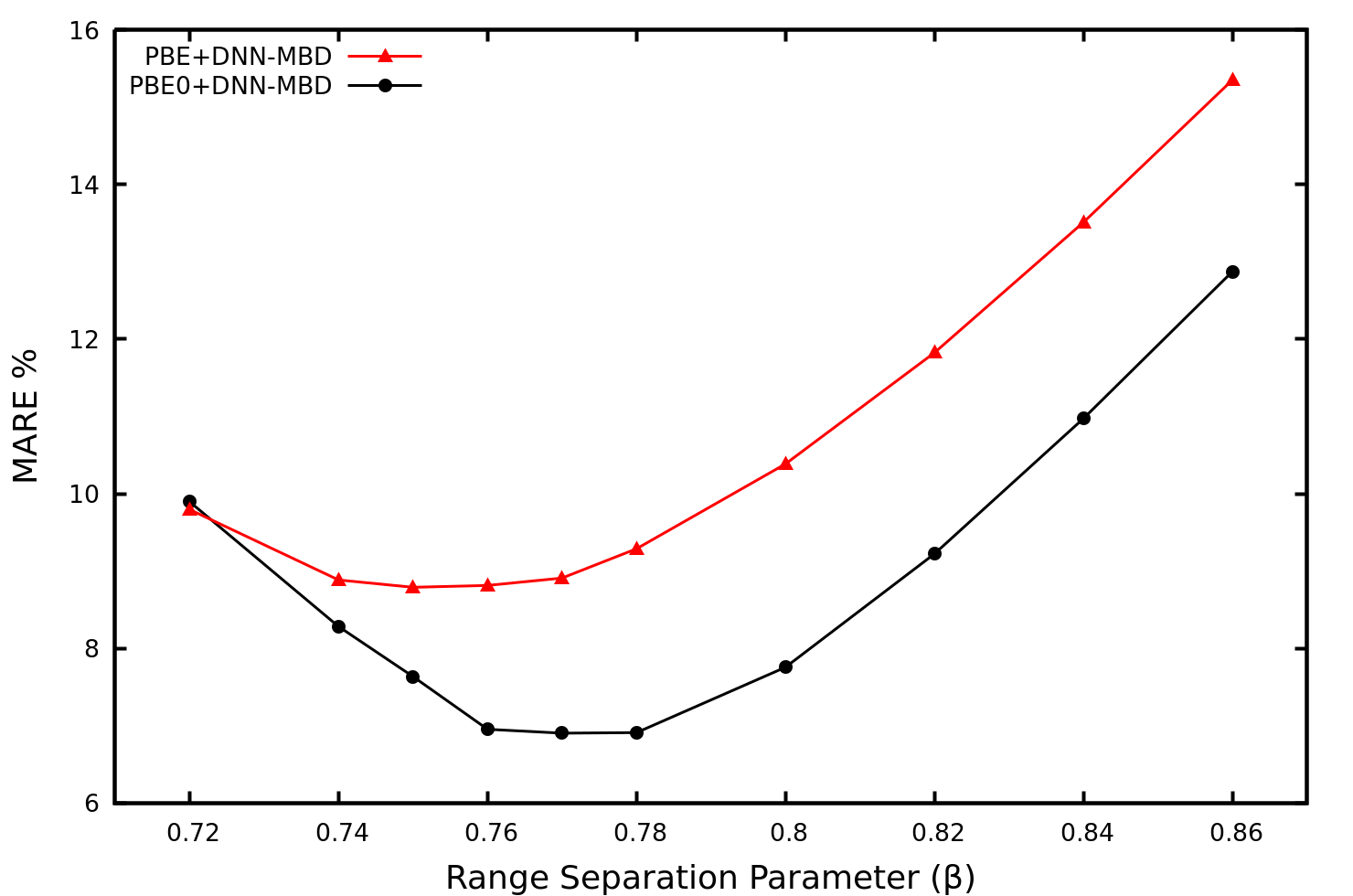}
   \caption{MARE (\%) as a function of the range separation parameter for the PBE+DNN-MBD and PBE0+DNN-MBD methods.}
   \label{fig:mare} 
\end{figure}

\begin{figure}[!htb]
\centering
   \includegraphics[width=1.0\linewidth]{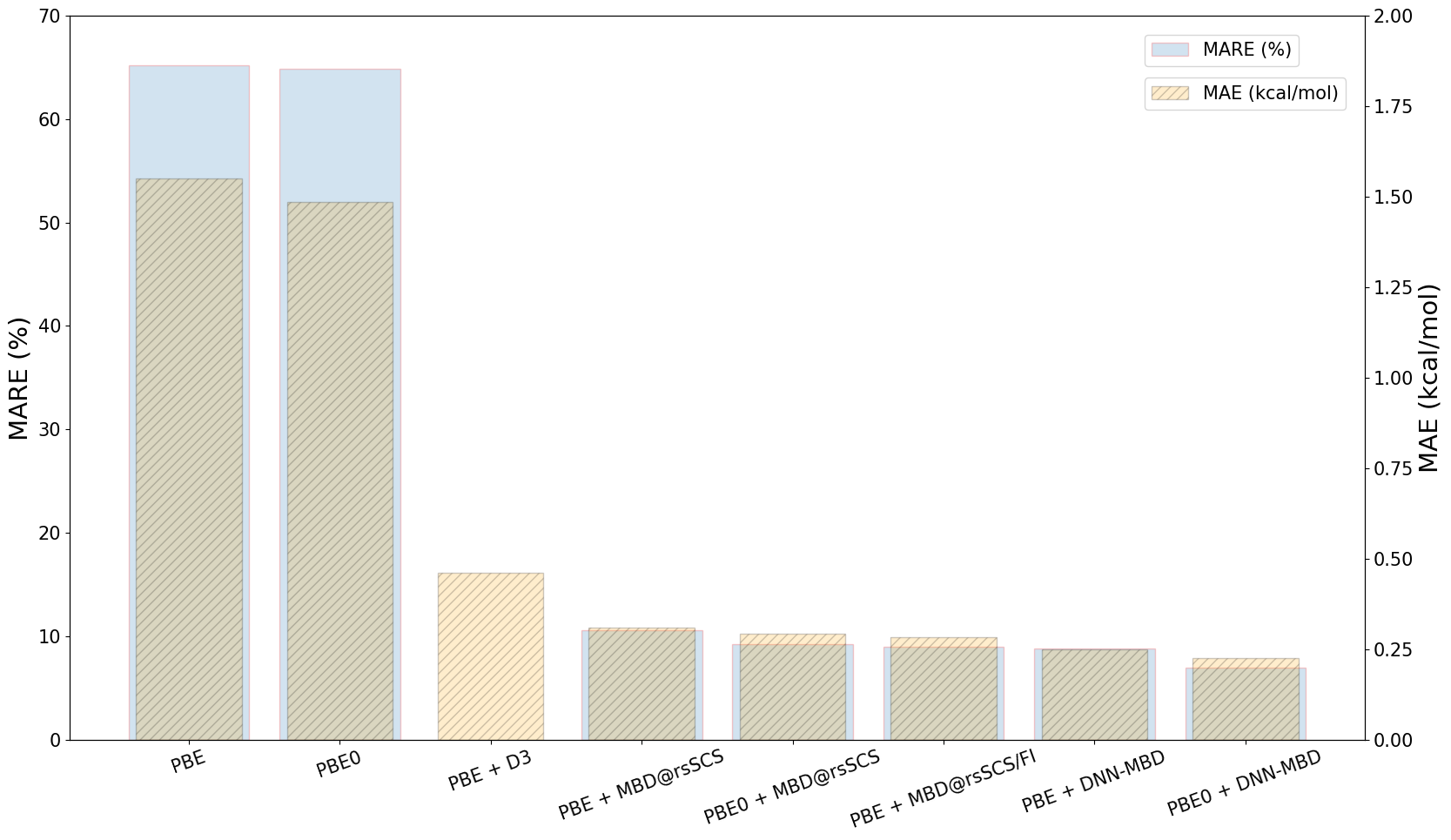}\par
   \includegraphics[width=1.0\linewidth]{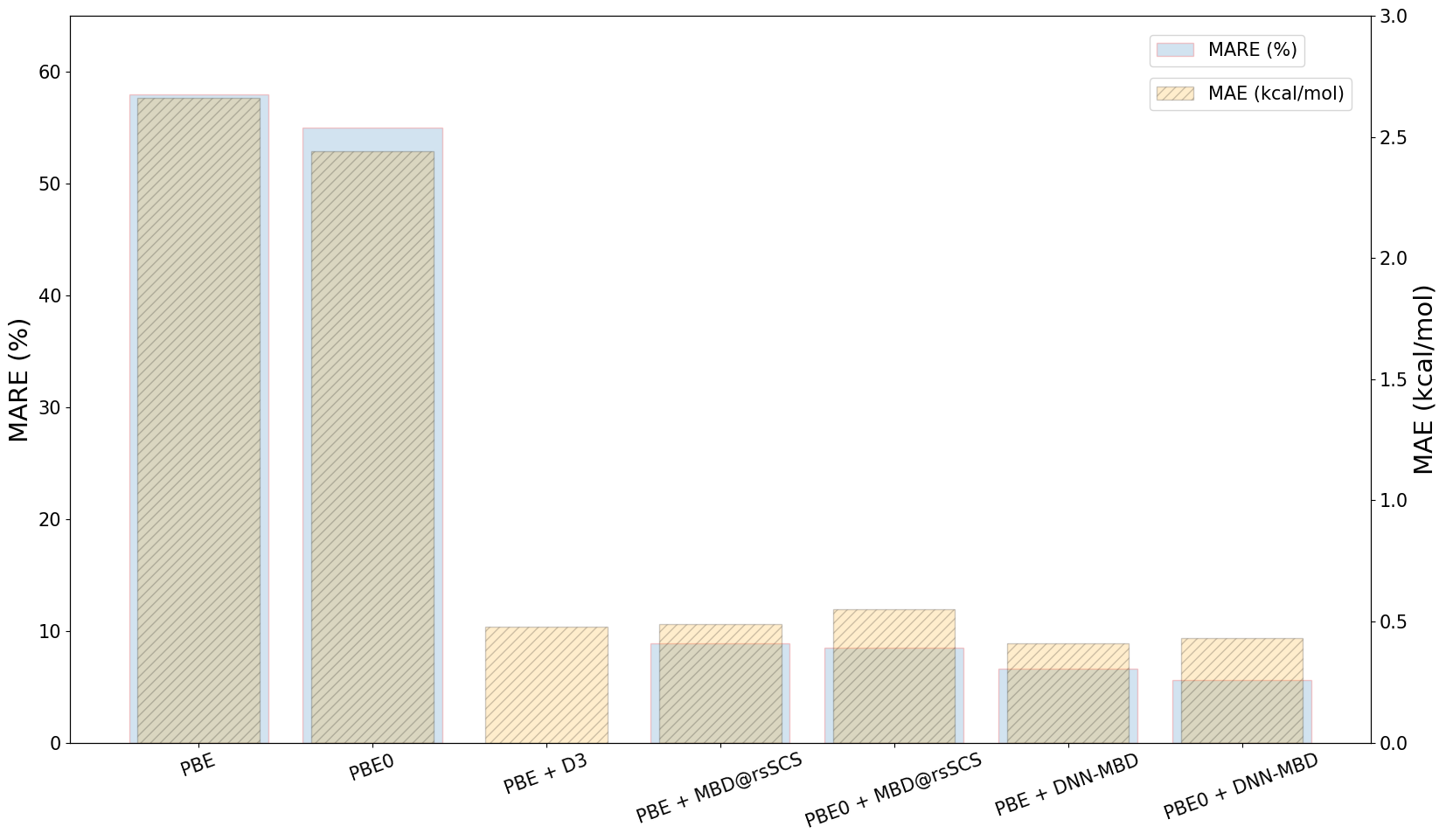}
   \caption{MARE (\%) and MAE (kcal/mol) of PBE, PBE0, PBE+D3\cite{grimmes66x8} and different MBD models (MBD@rsSCS\cite{mbd_rsscs}, MBD@rsSCS/FI\cite{fracpol}) including our DNN-MBD for the S66x8 (top) and S22 (bottom) data sets.}
   \label{fig:mare_mae} 
\end{figure}

The DNN model providing AIM volumes' ratios is embedded in the Tinker-HP package where our linear-scaling and embarrassingly parallel stochastic MBD@rsSCS is also implemented.\cite{mbd_stoch}\\
The outcoming DNN-MBD model is coupled to the common semi-local PBE\cite{pbe} functional as well as its hybrid PBE0 version\cite{pbe0} since this choices allow for comparisons with results ready available in literature. The optimal range-separation $\beta$ parameters for both the PBE+DNN-MBD and PBE0+DNN-MBD methods are obtained by minimizing the mean absolute relative error (MARE) on the widely employed S66x8 benchmark set consisting of 66 dimers placed at 8 different intermolecular distances for a total of 528 different structures where CCSD(T) interaction energies computed at CBS are used as reference. \\
All DFT computations employed Jensen's pcseg-3 basis set belonging to the family of segmented polarization-consistent\cite{pc} basis sets which, for DFT calculations, exhibits lower basis set errors than other gaussian basis sets as well as higher computational efficiency at given cardinal number as these basis sets were explicitly designed and optimized for DFT.\cite{pcseg}\\
Figure \ref{fig:mare} shows the MARE as a function of the range separation parameter for PBE+DNN-MBD and PBE0+DNN-MBD methods.\\
The optimal $\beta$ parameters are found to be 0.75 and 0.77 for the PBE+DNN-MBD and PBE0+DNN-MBD methods respectively. These values differ from the ones optimized for the original PBE/PBE0+MBD@rsSCS models\cite{mbd_rsscs} and this has to be addressed to the different partitioning scheme employed. As pointed out by Vestraelen et al.,\cite{mbisa} AIM densities computed via the  Hirshfeld or HI partitioning, exhibit asymmetries i.e. they are aspherical with too much density in the bonding region. This density accumulation, relatively far away from the atomic
nucleus, leads to larger values of radial moments, thus leading to larger AIM volumes compared to the ones obtained via the MBISA scheme (unaffected from this asymmetry artifact) for which less screening of volume-scaled AIM quantities (smaller $\beta$) is most likely to be needed.\cite{mbisa}\\
We observe, nevertheless, that both PBE0+DNN-MBD and PBE0+MBD@rsSCS methods require a larger $\beta$ parameter compared to their PBE corresponding models and this is consistent with the PBE0's improved description of short-range exchange-correlation effects due to the fraction of exact exchange included in the functional, as discussed in reference.\cite{mbd_rsscs}\\
The performance of the optimized PBE/PBE0+DNN-MBD methods is compared to different MBD models in terms of MAE and MARE for the S66x8 data set and the results are summarized in Figure \ref{fig:mare_mae} where actual values are reported in Table \ref{tab:s66x8}. \\
For the benchmark set here employed, the DNN-MBD model exhibits lower (although by a contained margin) errors both in its coupling to the PBE and PBE0 functionals compared to the standard MBD@rsSCS approach based on Hirshfeld AIM volumes as well the PBE+MBD@rsSCS/FI approach based on the fractionally ionic polarizabilities and HI AIM volume partitioning. For both the chosen functionals, the outcoming DNN-MBD model provides a mean absolute error in the S66x8 interaction energies which is below 0.25 kcal/mol compared to the reference CCSD(T) CBS golden standard.\\
To strengthen the analysis, we additionally computed the MAE and MARE for the S22 data set\cite{s22} by employing the range separation parameters previously optimized for the S66x8 set. We can, in this way, employ the S22 set as a test set to validate our conclusions, Figure \ref{fig:mare_mae} (bottom) and Table \ref{tab:s22}.\\
Compared to the S66x8 set, the MAE and MARE values of our proposed PBE/PBE0+DNN-MBD models are, for the S22 set, higher however this is not surprising as no $\beta$ optimization was performed this time. Let's note that all methods present errors that are larger in the case of the S22 set compared to S66x8 (see Table \ref{tab:s66x8} and Table \ref{tab:s22}). Indeed, there are reasons for that and we can stress that the dimers employed in the S22 set are placed at equilibrium while the S66x8 set includes out of equilibrium dimers. In our case, the DNN-MBD model trained on S66x8, appears less biased towards equilibrium structures. Overall, as one can see from Table \ref{tab:s22}, our DNN-MBD remains highly transferable and, with an error below 0.43 kcal/mol compared to the reference CCSD(T) CBS golden standard, outperforms previous S22 results obtained with others methods.
\begin{table}[!htb]
\begin{center}
 \begin{tabular}{||c c c||} 
 \hline
 \textbf{Model} & \textbf{MAE[kcal/mol]} & \textbf{MARE}\% \\ [0.5ex] 
  \hline\hline
    PBE  &   1.55    &    65   \\
    PBE0  &   1.48    &    65   \\
    PBE+D3  &   0.44    &    n.a.   \\
    PBE+MBD@rsSCS ($\beta=0.83$)  &   0.32    &    10.6   \\
    PBE0+MBD@rsSCS ($\beta=0.85$)  &   0.30    &    9.2   \\
    PBE+MBD@rsSCS/FI ($\beta=0.83$)  &   0.28    &    9.0   \\
    PBE+DNN-MBD ($\beta=0.75$)  &   0.25    &    9.0   \\
    PBE0+DNN-MBD ($\beta=0.77$) &   0.23    &    6.9   \\
\hline
\end{tabular}
\caption{\label{tab:s66x8}MAE (kcal/mol) and MARE(\%) relative to the S66x8 data set for our DNN-based models as well as for few other dispersion correction ones. For the MBD-based models, the method-specific range separation parameter reported in parentheses refers to the one optimised for the S66x6 set. MAE and MARE are computed taking revised CCSD(T) CBS energies.}
\end{center}
\end{table}

\begin{table}[!htb]
\begin{center}
 \begin{tabular}{||c c c||} 
 \hline
 \textbf{Model} & \textbf{MAE[kcal/mol]} & \textbf{MARE}\% \\ [0.5ex] 
  \hline\hline
    PBE  &   2.66    &    58   \\
    PBE0  &   2.44    &    55   \\
    PBE0+MBD@rsSCS ($\beta=0.85$)  &   0.55    &    8.5   \\
    PBE+MBD@rsSCS ($\beta=0.83$)  &   0.49    &    8.9   \\
    PBE+D3  &   0.48    &    n.a.   \\
    PBE0+DNN-MBD ($\beta=0.77$)  &   0.43    &    5.6   \\
    PBE+DNN-MBD ($\beta=0.75$) &   0.41    &    6.6   \\
\hline
\end{tabular}
\caption{\label{tab:s22}MAE (kcal/mol) and MARE(\%) relative to the S22 data set for our DNN-based models as well as for few other dispersion correction ones. For the MBD-based models, the method-specific range separation parameter reported in parentheses refers to the one optimised for the S66x6 set. MAE and MARE are computed taking revised S22 energies where, compared to the original paper, a larger basis set was employed.\cite{s22sherrill}}
\end{center}
\end{table}

Having been trained on a large and diverse set of AIM volumes, the outcoming DNN-MBD model inherits the strengths of the MBISA scheme discussed earlier in this Letter while completely bypassing the explicit density partitioning with a consequent decrease of the computational cost. We also note that the DNN model could be successfully trained with different AIM partitioning schemes due to the locality of the target quantities (volumes).\\ 
The presented density-free DNN-SMBD model is included in the Tinker-HP package\cite{tinkerhp} and will be released with the next version of the software. There, it can benefit from the linear-scaling embarrassingly parallel performances of our stochastic formulation (SMBD) of the MBD key equations which remarkable computational performances have been recently discussed.\cite{mbd_stoch} \\
We believe that the present DNN-SMBD model can be beneficial in applications of dispersion-corrected DFT to large complex systems requiring an accurate yet extremely efficient inclusion of MBD effects. The DNN model, by avoiding the direct solution of the KS equations due to its electron density-free features, allows for the ready application of DNN-SMBD approach in the development of accurate \emph{ab initio}-based force fields\cite{doi:10.1021/ct700134r,naseem2022development} and neural networks methodologies.

\begin{acknowledgement}
This work has received funding from the European Research Council (ERC) under the European Union’s Horizon 2020 research and innovation program (grant agreement No 810367), project EMC2 (JPP). Computations have been performed at GENCI (IDRIS, Orsay, France and TGCC, Bruyères le Chatel) on grant no A0070707671.

\end{acknowledgement}

\begin{suppinfo}
SI1 file contains training plots for DNN with different layers. SI2 and SI3 contain raw PBE, PBE+DNN-MBD, PBE0 and PBE0+DNN-MBD energies for the S66x8 and S22 data sets respectively.

\end{suppinfo}

\bibliography{achemso-demo}

\end{document}

\section{Data set volumes and volumes ratio.}
\begin{figure}[!htb]
\centering
   \includegraphics[width=0.7\linewidth]{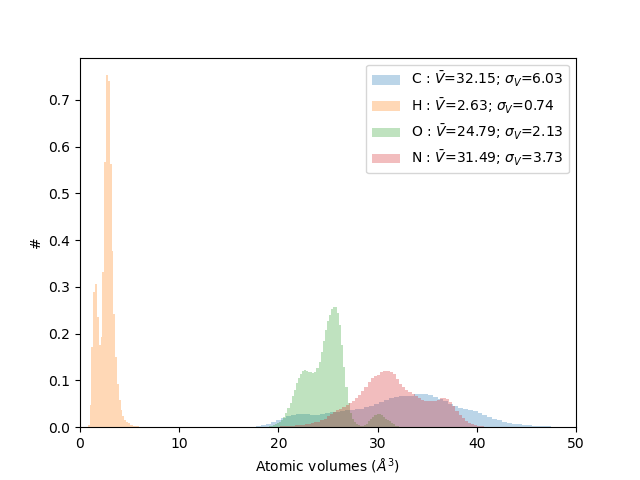}\par
   \includegraphics[width=0.7\linewidth]{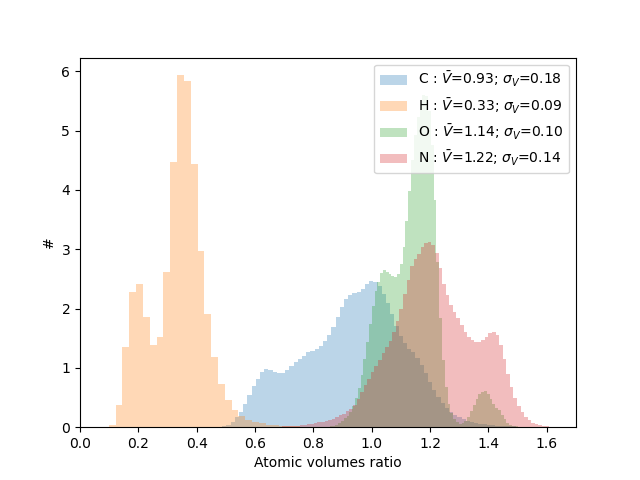}\par
   \caption{Atomic volume (top) and atomic volume ratio (bottom) normalized distributions of the ANI-1 dataset for the C, H, N, O atomic elements.}
   \label{fig:si_anidist} 
\end{figure}

\begin{figure}[!htb]
\centering
   \includegraphics[width=0.7\linewidth]{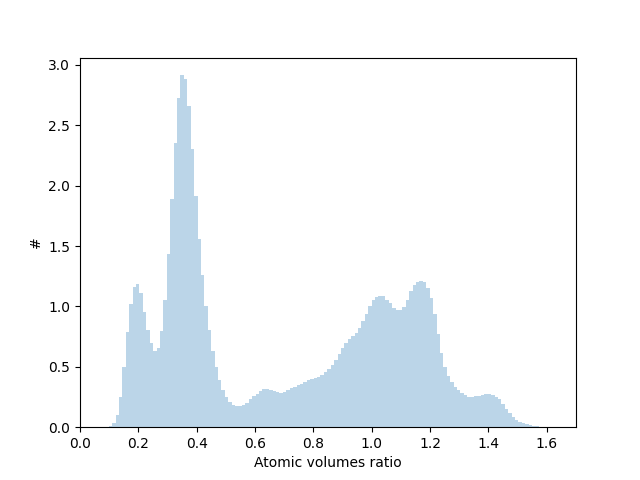}\par
   \includegraphics[width=0.7\linewidth]{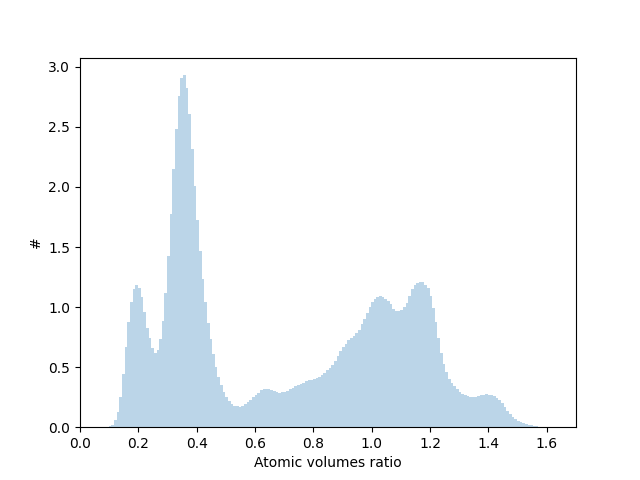}
   \caption{Atomic volume ratio distribution of the training (top) and the validation (bottom) set.}
   \label{fig:si_atomicdist} 
\end{figure}

Distribution of atomic volumes and atomic volumes ratio of the C, H, N and O atomic elements inside the ANI-1 dataset, composed of 4.6 millions of structures are shown in Figure \ref{fig:si_anidist}. Each atomic distribution is normalized to 1. Figure \ref{fig:si_atomicdist} shows the total atomic volumes ratio of the training and validation sets generated from the ANI-1 one. 

\section{Model training and validation errors.}
Below the RMSE evolution as a function of the epoch number is shown for both the training and the validation sets for the four different DNN architectures. In the case of the AIMNET-like model the learning rate reaches the early stopping learning rate within 3000 epoch. The final architectures employed for the final DNN-MBD model is the ANI-AEV 5 layers one.
\begin{figure}[!htb]
\centering
   \includegraphics[width=0.6\linewidth]{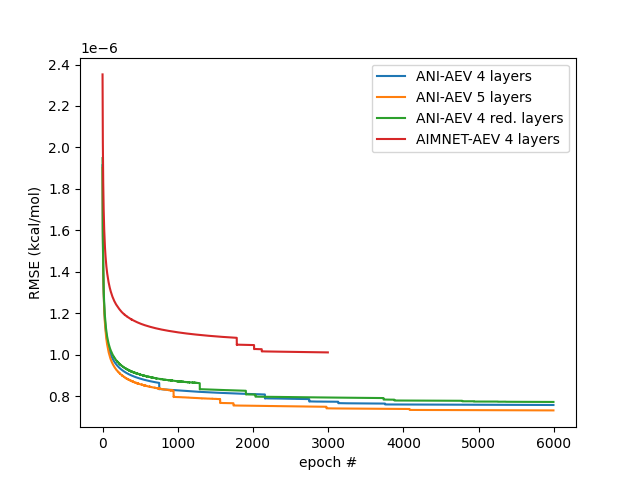}\par
   \includegraphics[width=0.6\linewidth]{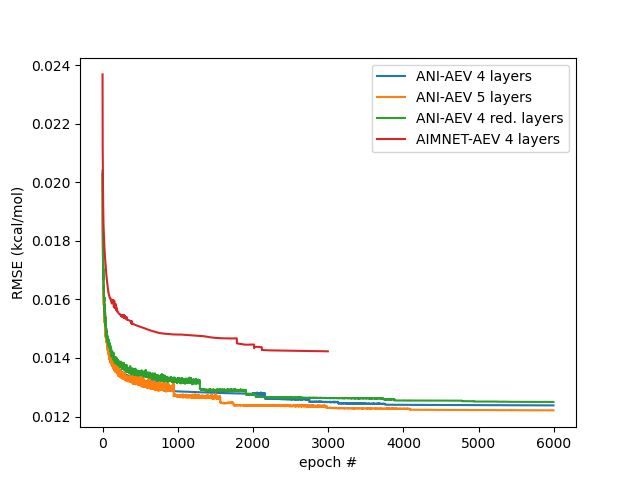}
   \caption{Evolution of the RMSE on volumes ratio w.r.t the number of epoch for the training (top) and the validation (bottom) set.}
   \label{fig:si_learningcurve} 
\end{figure}
The final DNN model and the dataset used for the training can be download directly via the Zenodo repository located at the following address : https://doi.org/10.5281/zenodo.6397639. \\
%\section{Interaction energy values for the S66x8 data set}
%In the provided \textbf{supplementary\_info\_energies.ods} file, the final interaction energies, the absolute dimer energies and isolated monomers' energies are given for the pure DFT models PBE/pcseg-3, PBE0/pcseg-3 as well as for the dispersion correction computed with the DNN-MBD model for a range of $\beta$ (range separation parameter) values.


%\bibliography{achemso-demo}